\documentclass{article}
\usepackage[latin1]{inputenc}
\pdfoutput=1
\usepackage{geometry}
\usepackage{mathtools}
\usepackage{amsmath}
\usepackage{amsthm}
\usepackage{amssymb}
\usepackage{cancel}
\usepackage{graphicx}
\usepackage[unicode=true]
 {hyperref}

\makeatletter
  \theoremstyle{definition}
  \newtheorem{defn}{\protect\definitionname}

\usepackage{multicol}
\usepackage{lscape}
\usepackage{iftex}
\ifPDFTeX
  \usepackage{microtype}
  \DisableLigatures[f]{encoding = *, family = * }
  \usepackage{accents}
\fi
\usepackage{setspace}
\usepackage{subcaption}
\usepackage{color}
\usepackage{sgame}
\usepackage{url}
\usepackage{wrapfig}
\usepackage{sgamevar}
\usepackage{booktabs}
\usepackage{amsthm}\usepackage{staves}\usepackage{enumitem}
\usepackage{titlesec}
\usepackage{tikz}
\usepackage{textcomp}
\usepackage{listings}
\ifPDFTeX
  \pdftrailerid{} 

\fi
\ifLuaTeX
  \pdfvariable suppressoptionalinfo 767

\fi
\usetikzlibrary{arrows}

\usepackage{nccmath}

\ifPDFTeX
  \renewcommand{\hat}{\widehat}
\fi

\ifPDFTeX
  \renewcommand{\hat}{\widehat}
\fi

\makeatother

\usepackage{listings}
  \providecommand{\definitionname}{Definition}

\begin{document}

\title{Pricing Engine:\\
 Estimating Causal Impacts in Real World Business Settings\thanks{We would like to thank Michael Schwarz, Preston McAfee, Matt Taddy,
Paul Chang, Aadharsh Kannan, Asaph Yeh, Aidan Marcuss, as well as
Microsoft's Algorithms and Data Science (ADS) and Market Intelligence
groups for valuable comments. We would also like to thank the MSR
Project ALICE expedition for their help in co-development of some
parts of the Pricing Engine code base. Contacts: mattgold@microsoft.com,
Brian.Quistorff@microsoft.com.}}

\author{Matt Goldman, Microsoft AI \& Research \\
Brian Quistorff, Microsoft AI \& Research}

\date{April 18, 2018}

\maketitle

\begin{abstract}
We introduce the Pricing Engine package to enable the use of Double
ML estimation techniques in general panel data settings. Customization
allows the user to specify first-stage models, first-stage featurization,
second stage treatment selection and second stage causal-modeling.
We also introduce a \texttt{DynamicDML} class that allows the user
to generate dynamic \textit{treatment-aware} forecasts at a range
of leads and to understand how the forecasts will vary as a function
of causally estimated treatment parameters. The Pricing Engine is
built on Python 3.5 and can be run on an Azure ML Workbench environment
with the addition of only a few Python packages. This note provides
high-level discussion of the Double ML method, describes the packages
intended use and includes an example Jupyter notebook demonstrating
application to some publicly available data. Installation of the package
and additional technical documentation is available at \href{https://github.com/bquistorff/pricingengine}{github.com/bquistorff/pricingengine}
.
\end{abstract}

\newpage{}

\begin{multicols*}{2}

\section{Introduction}

The explosion of data science in modern technology firms has created
a new class of workers with the technical backgrounds needed to solve
a wide array of statistical problems using a diverse set of machine
learning (ML) techniques. However, the most important decisions made
by such firms are typically \textit{policy questions} such as \textit{How
much should we invest in R\&D?}, \textit{Should we cut prices?}, or
\textit{Which product would benefit most from an aggressive marketing
campaign?}. These are all questions that hinge on understanding the
\textit{causal} effect of various policy interventions and, as such,
cannot be answered (or even well-informed) by purely statistical approaches.
Instead, they require econometric techniques that can yield answers
with a clear \textit{causal} interpretation.

Causal inference is about understanding the true effect of a treatment,
call it `$D$', on an outcome, call it `$Y$'. How would $Y$ change
if we changed D? ML on the other hand is usually about building a
good predictor function of $Y$ using many features $X$ (that may
include $D$). These are fundamentally different and therefore one
should be careful when moving from one domain to the other or combining
the two.

When $D$ is randomly assigned, establishing a causal interpretation
is straightforward and can be done using even very simple statistical
techniques. However, most business decisions involve situations where
experimental data is impossible to attain or at least not immediately
unavailable. This introduces considerable additional difficulty. One
has to control for factors that independently affect both $Y$ and
$D$. These are called confounders (call them $X$). Omitting these
will cause our causal estimates to be wrong. This is often referred
to as ``omitted variable bias''. As a simple example, suppose we
were to analyze the impact of price on sales volume for a retailer
of children's toys, but we were unaware of the important role of the
holiday season. We might naively conclude that the relatively small
discounts observed at Christmas time were very effective drivers of
sales and we might erroneously advise the retailer to consider lowering
it's prices more often. Should they follow our advice and Christmas
should ``fail to come in July'', our error could be quite embarrassing.

Even a naive analyst could be expected to recognize the confounding
role of Christmas, but this is an extreme case. Often confounding
variables are much more subtle and an analyst may be unsure whether
a particular variable should be considered a confounder or how it
interacts with other elements of the system. This process, of carefully
selecting and modeling the impact of confounds, is referred to by
economists as \textit{model selection}. In practice, this is often
a time-consuming, tenuous and somewhat arbitrary process that can
make a significant difference in the final result. Economist's are
often loathe to trust empirical work done by non-experts because they
will feel that this \textit{model selection} stage has been performed
in error.

It is this last issue \textendash{} the difficulty of model selection
\textendash{} that is the motivation of our package. We seek to automate
the process of model selection by placing machine learning techniques
within a carefully constructed econometric framework that can deliver
robust causal estimates. For economists this offers the attraction
of automating model selection to generate significantly improved (and
less arbitrary) models. For machine learning practitioners, this has
the advantage of placing the necessary structure around their tools
to ensure a clean causal interpretation of business-relevant parameters.
For a business person, this can yield an all in-one solution that
generates \textit{treatment-aware} forecasts, providing forward looking
predictions for future outcomes of interest (e.g. sales) and how (on
average) those outcomes can be modified as a function of planned treatments
(e.g. prices).

\section{Double ML Preliminaries\label{sec:Double-ML-Preliminaries}}

The econometric framework that allows us to use ML in causal inference
is referred to as ``Double ML'' \cite{chernozhukov2016double}.
Its application to the problem of controlling for confounders rests
on two key ideas: the Frisch\textendash Waugh\textendash Lovell (FWL)
Theorem and the ``cross-fitting'' procedure. The FWL theorem is
a simple application of linear algebra to ordinary least squares (OLS)
regression. Suppose we want to estimate the effect of $D$ on $Y$
while controlling for $X$ in a regression without any ML. The standard
way would be to estimate the full regression model:
\begin{align}
Y & =\beta\cdot D+\gamma\cdot X+\epsilon\label{eq:Y_linear}
\end{align}
The FWL Theorem states that we will recover the same estimate of $\beta$
by estimating equation~\ref{eq:Y_linear} as if we first remove the
effect of $X$ from both $Y$ and $D$. That is if:
\begin{enumerate}
\item Regress $Y$ on $X$. Then generate fitted values $\hat{Y}$ and
residuals $\tilde{Y}=Y-\hat{Y}$.
\item Regress $D$ on $X$. Then generate fitted values $\hat{D}$ and
residuals $\tilde{D}=D-\hat{D}$.
\item Estimate $\tilde{Y}=\beta\cdot\tilde{D}+\epsilon$
\end{enumerate}
The first two steps can be thought of as ``baseline'' estimates
of $Y$ and $D$ that just use $X$. Notice that we do not care about
the coefficients in the baseline stage. All we care about is how to
predict the outcome and treatments as a function of the potential
confounders. However, using OLS to fit the regressions in steps (1-2)
has two key weakness: first, it does not allow for a non-linear relationship
between confounders and outcomes/treatments and second, if the number
of confounders is not small (when compared to the sample size), OLS
will substantially overfit the model in a way that has poor out of
sample predictive properties.

As such, we may prefer to use more general ML techniques to fit the
baseline estimations. However, without further accommodation, overfitting
can still lead to poor statistical performance even if our ML algorithm
has strong predictive performance. Overfitting, though, is not a new
problem for the ML literature. The key then is to re-purpose the existing
solution of cross-validation into a new algorithm called ``cross-fitting''. 
\begin{defn}
{[}$K$-fold Cross-fitting{]} \label{def:cross-fitting} Cross-fitting
is a procedure for fitting and predicting a model type $f$ on data
using multiple sub-models $\{f_{1},...,f_{K}\}$ of the same type
in such a way that predictions for each observation are done using
sub-models that were not trained on that observation. 
\begin{enumerate}
\item Fitting: 
\begin{enumerate}
\item Split the data into a $K$-fold partition. 
\item For each partition $k$, fit $\hat{f}_{k}$ by excluding the data
from partition $k$.
\end{enumerate}
\item Prediction: For each observation $i$ with features $x_{i}$, the
prediction $\hat{f}$ is the average prediction of all sub-models
that were not trained on observation $i$. 
\begin{itemize}
\item If observation $i$ was used in fitting $\hat{f}$, then $\hat{f}(x_{i})=\hat{f}_{k}(x_{i})$
for the $k$th fold that contained observation $i$. 
\item If observation $i$ was not used in fitting $\hat{f}$ (e.g. we are
looking at a hold-out sample), then $\hat{f}(x_{i})=\sum_{k=1}^{K}\hat{f}_{k}(x_{i})$.
\end{itemize}
\end{enumerate}
\end{defn}
Cross-fitting can be thought of as the first phase of cross-validation.
Cross-validation would normally continue and look at average predictive
performance on the test folds, for example, as part of a larger algorithm
to tune a hyper-parameter.

By using cross-fitting in our baseline predictive stages, our third
regression is performed on residuals that are calculated using models
trained on entirely independent data. These are called ``honest''
residuals and have much better statistical properties.

We are now ready to pose the full ML problem. This involves structuring
the overall causal inference question so as to split out parts that
are pure prediction problems and as such can be hand over to ML. To
be concrete, imagine that we are interested in some parameter $\beta$
which gives the average effect of treatment ($D$) onto an outcome
($Y$). Further, suppose that we observe a high-dimensional set of
\textit{potential confounds} ($X$). Each element of $X$ might be
related to the outcome, to the assignment of treatment, or both. As
such we will not be able to learn $\beta$ until we have modeled the
impact of these confounding variables. Suppose we place absolutely
no restrictions on the impact of our confounders, then we would write
\begin{align}
Y & =\beta\cdot D+f(X)+\epsilon\\
D & =g(X)+\mu\\
E & [\epsilon\cdot D|X]=0,\label{eqn:orthog}
\end{align}
where $f$ and $g$ are unrestricted functions. These are nuisance
parameters of the system, and let their collection be $\eta=(f,g)$.
One important note, is that \eqref{eqn:orthog} has imposed an important
restriction on our model. Often referred to as an \textit{independence}
or \textit{orthogonality} condition, this implies that once we have
correctly modeled the impact of our confounding variables on our treatment,
any remaining variation in treatment must be uncorrelated with the
outcome. If $Y$ was sales and $D$ was price, this condition would
require that idiosyncratic variation in sales (over and above what
we could forecast in our first regression) was not driven by the same
unobservables that drove idiosyncratic variation in prices. If there
are major demand shocks of which price-setters were aware, but which
are not recorded in our data, then we may find this assumption to
be questionable and may need to consider other approaches. But, if
we accept this condition (as we might if $X$ contains all the important
data available to our price-setters), then we can perform robust causal
inference on $\beta$ by using the following steps. First, ignore
treatment and estimate a reduced form relationship ($l$) between
$X$ and our outcome $Y$.\footnote{Note that when we are estimating the reduced form relationship between
$Y$ and $X$, we do not recover the structural impact of $X$ on
$Y$ (as given by $f$), but rather the related object $l$ which
also includes the downstream impact our confounders have on the outcome
that is channeled through treatment. }
\begin{align*}
l(X) & \equiv E[Y|X]=f(X)+\beta\cdot g(X)
\end{align*}
 Second, estimate $g$ as the reduced form relationship between our
confounds X and our treatment $D$. These two steps are analogous
to ``baseline'' (or treatment-blind) forecasts of our both our outcome
and our expected assignment of treatment. These steps are analogous
to steps 1 and 2 in the FWL Theorem and we refer to them collectively
as \textit{first-stage regressions}. Arbitrary ML methods can be used
to estimate these objects with the maximum possible out-of-sample
precision. The residuals from these regressions can then be considered
to be ``surprises'' in the evolution of treatment/outcome and any
remaining correlation between them can then be estimated, in a \textit{second-stage
regression} (analogous to the final step of the FWL Theorem) and interpreted
as robust evidence of a causal effect of treatment on outcome. We
formalize this recipe (and also describe how cross-fitting is used)
in \ref{def:dml_recipe} below.
\begin{defn}
{[}Double ML Recipe{]} \label{def:dml_recipe} The generic Double
ML recipe is: 
\begin{enumerate}
\item Split the data into a $K$-fold partition. 
\item Estimate $\hat{l}$ and $\hat{g}$ by cross-fitting using the common
data partition. 
\item Compute first-stage residuals. 
\begin{align*}
\tilde{Y} & =Y-\hat{l}(X)\\
\tilde{D} & =D-\hat{g}(X)
\end{align*}
\item Pool the first-stage residuals from all partitions and estimate the
causal effect ($\hat{\beta}$) by a simple regression of $\tilde{Y}$
onto $\tilde{D}$. Additionally, the OLS standard errors computed
from this regression can be interpreted as a valid standard error
for $\beta.$  
\end{enumerate}
\end{defn}
This recipe was originally proposed by Chernozhukov et al. \cite{chernozhukov2016double}
and proven to be consistent for inference on a low-dimensional treatment
parameter $\beta$. It was extended to the case where treatment is
high dimensional (i.e. we may want to learn a full matrix of cross-price
elasticities) \cite{chernozhukov2017orthogonal}.

\subsection{Extension to heterogeneous treatment effects}

Suppose we are interested in understanding how the impact of $D$
varies with some other co-variate. For example, we might want to test
the hypothesis that consumer demand responds to some treatment (perhaps
a price cut) by more in the US as compared to other markets. Thus
we might modify our previous model to take the slightly altered form
given by

\begin{align}
Y & =\beta\cdot D+\beta_{2}\cdot D\cdot1\{Market=US\}+f(X)+\epsilon\\
D & =g(X)+\mu\\
E & [\epsilon\cdot D|X]=0.\label{eqn:orthog-1}
\end{align}
Here it is sufficient to follow the same algorithm as in \ref{def:dml_recipe},
but to alter the final stage by regression $\tilde{Y}$ jointly onto
$\tilde{D}$ and $\tilde{D}\cdot1\{Market=US\}$ to learn both a baseline
treatment effect and a heterogeneous impact of treatment in the US
Market. More generally, shows that this algorithm may be used to learn
any set of heterogeneous treatment effects which are an affine modification
of a single core treatment which is residualized \cite{chernozhukov2017orthogonal}.
In addition to the simple interaction demonstrated above, this can
include higher order interactions, the impact of a peer's treatment,
or the average impact of peer treatment averaged over a broad set
of peers (e.g. an average cross-price elasticity over some range of
competing products). However, this method cannot be applied to learn
the impacts of non-linear transformations of the core-treatment (e.g.
$D^{2})$. In such cases we must preform an entirely separate residualization
as demonstrated in the next subsection. 

\subsection{Extension to multiple treatments}

Often we want to estimate the impact of multiple treatments. For example,
we may wish to model sales as a function of both pricing and marketing
treatments. Often pricing and marketing decisions are correlated (it
may make sense to run a price cut contemporaneously with a big ad
purchase) and as such we must model these two treatments simultaneously.
To do otherwise \textendash{} estimating their effects separately
\textendash{} would attribute the impacts of both treatments to whichever
one was being modeled. More generally, suppose we want to estimate
$P$ different treatment effects. Let our environment be 
\begin{align}
Y & =f(X)+\sum_{p\in P}\beta_{p}\cdot D_{p}+\epsilon\\
D_{p} & =g_{p}(X)+\mu\qquad\forall p\in P\\
E & [\epsilon\cdot D_{p}|X]=0\qquad\forall p\in P.\label{eqn:orthog_multi}
\end{align}
Then our previous procedure is modified so that we train and independent
predictive function $g_{p}$for each treatment $D_{p}$. Then, following
the established cross-fitting formula, we compute residuals $\tilde{D}_{p}=D_{p}-\hat{g}_{p}(X)~\forall p\in P$
and then jointly regress these residuals onto $Y.$

\section{Dynamic DML}

On it's own, Double ML doesn't incorporate any explicit knowledge
of how data or effects are related across time. But in a business
context, this can be very important. There is often a significant
gap between when actions need to be planned and when they can be executed,
but the size of this gap can vary dramatically across contexts. Supply
chain decisions often require an initial purchase to be executed months
in advance, marketing budgets can be adjusted weeks in advance, and
in some cases pricing can be adjusted at a mere moments notice. Furthermore,
we may often want to understand the impact of a price chosen tomorrow
on consumer demand into the future (did we cannibalize some future
demand?). These concerns lead us to develop the Dynamic DML algorithm,
which is an extension of Double ML to a setting where we must model
outcomes a variety of lead times.

Dynamic DML incorporates this into both the baseline forecasting and
causal model stages. At the baseline stage, we need not just one baseline
forecasting model ($\hat{l}$, $\hat{g}$), but rather a range of
forecasting models for different lead times. For example, a forecast
with a lead of one would mean a forecast that is looking one period
ahead. Formally, for each lead $\tau$, we want to train baseline
models over data from each unit $i$ and time period $t$:
\begin{align*}
Y_{i,t+\tau} & =l_{\tau}(I_{it})\qquad\forall i,t\\
D_{i,t+\tau} & =g_{\tau}(I_{it})\qquad\forall i,t,
\end{align*}
where $I_{it}$ is all the information known at time $t$ such as
$X_{it}$, $D_{it}$, $Y_{it}$, their past values, and anything predetermined
(e.g. dates of holidays). The analogue is of a forecaster at some
\emph{reference date} trying to predict values at some future \emph{outcome
date} and the \emph{lead} \emph{time} is the difference between the
two. Each variable's lead time-specific forecast is estimated across
all possible values of the reference date. We formalize this view
in the following definitions.
\begin{defn}
Time-frames:
\begin{itemize}
\item \textbf{Reference date \textendash{}} the date from which we sit when
we train our first-stage baseline models to predict outcome and treatment. 
\item \textbf{Outcome date} \textendash{} the date for which the forecaster
wants to predict outcomes. 
\item \textbf{Lead time} \textendash{} the gap between the \textbf{outcome
date} and the \textbf{reference date}. 
\end{itemize}
\end{defn}
With multiple forecasts at different leads we can extend the causal
model to identify delayed effects of some treatment (e.g. the ``pull-forward''
effect of a sale). Suppose we trained forecast models for leads one
to four. The residuals $\tilde{D}_{i,t+1}$ and $\tilde{Y}_{i,t+1}$
are the one-period-ahead surprises in treatment and outcome. The relation
between them gives us evidence for the contemporaneous treatment effect.
With more leads, we can go further and look at how $\tilde{D}_{i,t+1}$
affects $\tilde{Y}_{i,t+2}$, which helps us identify a delayed treatment
effect. In economic terms, this gives inter-temporal substitution
(or the ``pull forward'' cannibalization of demand). Figure~\ref{fig:dynamic_dml_leads}
shows this graphically for this example setup and think of the forecaster
at date $t_{0}$.

\includegraphics[width=1\columnwidth]{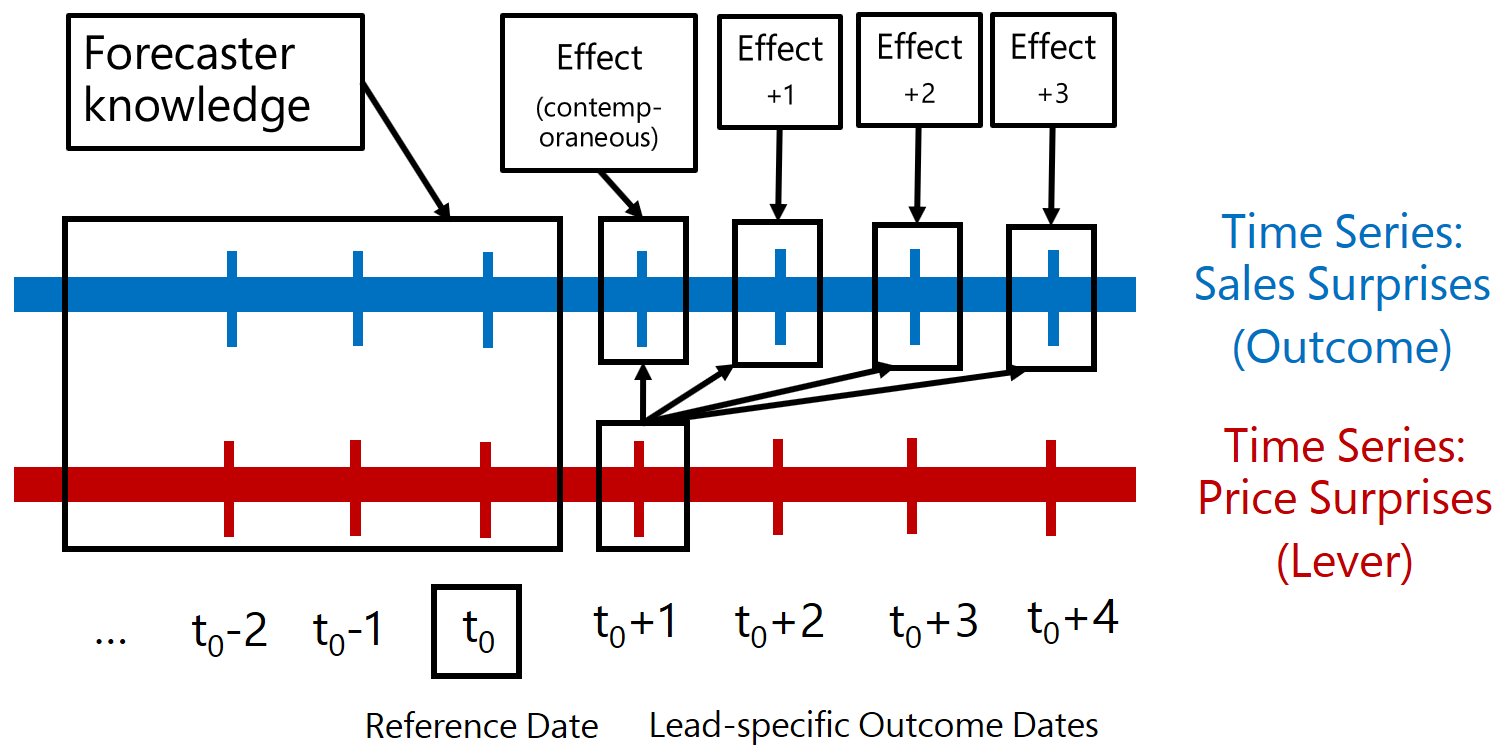}

\captionof{figure}{How DynamicDML structures forecasting and estimation} 

\label{fig:dynamic_dml_leads}

The \texttt{DynamicDML} object takes a set of leads, builds the accompanying
baseline forecasts, and then allows the user to estimate intertemporal
effects. If one wishes to ignore these dynamic complexities, the \texttt{DoubleML}
class (which defaults to a single lead of zero) may offer a simpler
user experience. Or one can simply set \texttt{min\_lead=max\_lead=1}
in \texttt{DynamicDML} which trains a single model using the maximum
available information set (which typically results in the narrowest
confidence intervals on the resulting causal parameters).

\section{Modeling Recommendations}

\subsection{Baseline Stage\label{subsec:Baseline-Featurization}}

In this section we recommend what should be included in the base set
of features $X$ and how to construct the unknown $l$ and $g$ functions.
This latter part entails both the feature generation and picking the
ML algorithm, which are closely related (e.g. tree models will automatically
try to detect interactions and non-linearities whereas a Lasso will
need these explicit featurized).

In general it is better to add anything that might be a confounder
or predictor to $X$, but technically we must avoid what the Econometrics
literature calls \textit{bad controls}. These are variables whose
values were affected by the value of treatment. They are, therefore,
a type of ancillary outcome rather than a real control and their inclusion
in $X$ will bias our estimated treatment effects (they capture part
of the overall effect of $D$ on $Y$). In most settings, however,
the forecaster perspective used by Dynamic DML will be sufficient
to avoid the problem as only previous values of variables are used
as features in the baseline stage.

For feature generation, we provide the following built-in featurizers:
\begin{itemize}
\item \texttt{default\_panel\_featurizer}: Includes $X$, dummy variables
for each unit of time, and dummy variable for each panel/unit variable
(e.g. ``product''). This feature attempts to mimic standard ``panel
data'' analysis.
\item \texttt{default\_dynamic\_featurizer}: Includes $X$, past values
of $D$ and $Y$, and trends of those variables. This featurizer attempts
to detect trends in product popularity. 
\end{itemize}
In addition, both of these features can absorb additional features
of the reference date that may help us make forward looking predictions
(e.g. how many google searches do we observe for our product) or features
of the outcome date that may help predict seasonal patterns (e.g.
is the outcome date Christmas? What is the usual intensity of sales
during that time of year?). 

For ML algorithms, the following are some general recommendations
from Chernozhukov et al. \cite{chernozhukov2016double}:
\begin{enumerate}
\item If the set of true confounders is sparse (i.e. only a few are truly
important), use sparsity-based techniques such as Lasso, post-Lasso\footnote{Where we first run a Lasso to get selected variables and then run
OLS with just the selected variables.}, or $l_{2}$-boosting.
\item If confounders have sharply different behavior on different subsets
of our date, it may be best to use trees or random forests.
\item If $f$ and $g$ are well approximated by a sparse (deep) neural net,
then use an $l_{1}$-penalized (deep) neural network.
\item If any of the above are true, then one can also use an ensemble method
over the methods methods mentioned above.
\end{enumerate}
Following these guidelines, the Pricing Engine has built-in models
for Lasso, boosted trees, Random Forests, Neural Nets, as well as
simple ensemble methods such as Bucket of Models and Stacking. Alternatively,
the pricing engine can also take (as inputs to a \texttt{PrePredicted}
class) first-stage forecasts generated by some other ML tool.

\subsection{Causal algorithm\label{subsec:Causal-algorithm}}

The Double ML theory provides statistical guarantees for using OLS
as the causal algorithm and this should be the default choice for
most problems. OLS tends to perform badly, however, with many, highly-colinear
features. In these cases Ridge regression may provide more stable
second-stage estimates. The basic Ridge algorithm, however, has the
downside that, since it penalizes its parameters, the estimates will
be biased towards zero. We therefore provide a modified Ridge algorithm
that in practice captures the majority of the benefits of Ridge regression
while retaining only a fraction of its downsides. The key is that
usually the treatment effects can be clustered so that each group
contains a high-level, main effect and then numerous secondary effects.
For example, we may look for the average effect of a price discount
and then check for heterogeneous effects by each sales region. Given
we are checking for heterogeneous effects by all regions we should
be more skeptical about each of those than about the main effect.
Therefore we provide a modified Ridge algorithm that allows certain
features to be unpenalized and pair that with \texttt{TreatmentBuilder}
objects that can be configured to penalize just the ``secondary''
treatments effects.

\subsection{Diagnostics\label{subsec:Diagnostics}}

Evaluating the performance of a model's predictive ability is usually
straightforward: retain a hold-out ``test/validation'' sample and
check the fitted model's performance on that sample. This will give
an unbiased estimate of the true predictive performance. Evaluating
the validity of a model's causal estimates is much more difficult.
In observational data, there is usually no ``ground truth'' that
cleanly indicates true causal effects.\footnote{One can know the ground truth by generating artificial data with known
treatment parameters and seeing how well the estimation strategy does
at recovering the effects. Since this type of validation is quite
narrow (it only gives assurance about the particular data generating
process specified, not necessarily what is going on in the real world)
it may be best when there is a particular concern about the estimation
setting and the model choices under consideration.} This problem is not new, and there is an extensive literature on
validating and interpreting the output of OLS type methods. We briefly
mention here some of the key points and note how these methods can
be applied to the Double ML setup.

First we note that estimated coefficients may occasionally have an
unintuitive sign. As an example, if treatment is price and our outcome
is sales, a positive coefficient would indicate that a price increase
would result in more sales. This is not necessarily cause for concern.
A user should first check their estimated standard errors, to see
if their result is statistically significant. If the estimate is statistically
insignificant, it may be the case that there was simply too little
residual variation in your treatments in order to receive useful causal
estimates. If however, coefficients are estimated ``with the wrong
sign'' and are statistically significant, this may be cause to re-evaluate:
(1) the first-stage models and whether or not they are appropriately
specified to model the impacts of confounding variables or (2) the
independence assumption discussed in section~\ref{sec:Double-ML-Preliminaries}.

Second, we may want to see how robust our estimates are to small changes
in our estimation strategy. Here we suggest diagnostics that analyze
how model estimates change as either variables or observations are
dropped. One of the key metrics at the variable level is the \emph{Variable
inflation index (VIF)}. This is a matrix of the correlations between
the treatment variables. Highly correlated treatments will causes
the estimate of one parameter to depend on the inclusion of the other.
This will also mean that the two treatments will ``compete'' for
the effect often causing one to take an unintuitive sign. A Ridge
regression in the causal stage can be used to deal with highly correlated
treatments. A good baseline stage should also lower the VIF by projecting
out common drivers of both treatments so that effects are easier to
estimate. One may also be concerned that outlier observations are
driving their causal estimates. It may be wise to individually inspect
those observations who's values of treatment and outcome are fit the
most poorly in the first-stage regressions and to see how omitting
these data points impacts model estimates.\footnote{There are multiple statistics at the observation level that characterize
how the model changes with the inclusion of each observation. General
outlier analysis at baseline is helpful for determining problems in
the underlying data. After baseline one can look at outliers (in terms
of the outcome or treatment variables) to assess how well the model
is at finding the ``mini-experiments''. Are the large residuals
periods where we think there was some new change to the variable or
is there some confounding that the model is picking up? This can be
extended to looking at specific measures of ``influence'', such
as \emph{Cook's D} (how the overall fit of a regression changes with
the inclusion of each point) and \emph{DFBETA} (how the coefficients
change with the inclusion of a point) in the causal stage.}

Finally, even if our causal estimates are of reasonable sign and magnitude
and are not overly sensitive to outliers/model specification, we will
still want to validate them. Fundamentally, this can only be done
by randomizing the value of treatment (i.e. preforming an experiment).
In some cases, experimental variation may already be present in some
subset of the data. Then one can compare estimates on this subset
versus on its compliment. Alternatively, one can use the model's estimates
as a suggestion of where to target new experiments and in the process
validate the model estimates. 

\section{Implementing Double ML in the Pricing Engine SDK\label{sec:Implementing-Double-ML}}

The Pricing Engine SDK enables the user to flexibly apply this structure
to their problem of choice. The major choices available to the user
are:
\begin{enumerate}
\item Given the data, what features $X$ compromise the high-dimensional
set of potential confounds? Do we want to construct derived features
to capture particular dynamics, interactions, or non-linearities?
(See section~\ref{subsec:Baseline-Featurization})
\item What ML algorithm should be use for the \textit{first-stage regressions}
of $D$ and $Y$ onto $X$? (See section~\ref{subsec:Baseline-Featurization})
\item Exactly how residualized treatments should be manipulated so that
we can learn interesting patterns of heterogeneous, or peer, treatment
effects. This will depend on the stakeholders
\item What second-stage algorithm should be used to infer the causal parameter
$\beta$? (See section~\ref{subsec:Causal-algorithm} )
\item How far into the future do we wish to forecast outcomes and understand
causal impacts? 
\end{enumerate}
We encourage the reader to review the accompanying OJ Demand Model
Jupyter notebook to see an example of how these choices are specified.
A relevant code snippet is produced below. As you can see the user
can flexibly enter different values for 
\begin{enumerate}
\item \texttt{feature\_builders}: This is a list of \texttt{VarBuilder}
structures specifying how first-stage features are generated. In this
example, we use a default list of \texttt{VarBuilders} created by
another class, but this can just as easily be flexibly specified by
the user to contain a preferred set of forecasting features. You can
swap in your own featurizer function in place of \texttt{default\_panel\_featurizer}.
\item \texttt{baseline\_model}: The model used to estimate first-stage (ML)
regressions. New base models can added by inheriting from our \texttt{Model}
class (which will be automatically wrapped for cross-fitting) and
new ensembles methods can be added by inheriting from our \texttt{SampleSplitModel}
class. If you would like to generate the predictions offline, potentially
in a completely different environment, we provide the \texttt{PrePredicted}
class with utilities for integrating those predictions into \texttt{DynamicDML}.
\item \texttt{treatment\_builders}: A list of \texttt{VarBuilder} objects
specifying how residualized treatments will be modified before second-stage
regression. In this case we have used the \texttt{interaction\_levels}
parameter to get heterogeneous effects across a number of dimensions
and used the \texttt{PToPVar} class to specify peer (cross-price)
treatment effects. 
\item \texttt{causal\_model}: The second-stage regression model used to
learn the causal effects. 
\item \texttt{Options}: Where we have passed \texttt{min\_lead} and \texttt{max\_lead}
which govern the range of leads for which we preform baseline forecasting. 
\end{enumerate}
\begin{lstlisting}[basicstyle={\small\ttfamily}]
col = [ColDef('store', DataType.CATEGORICAL),
       ColDef('brand', DataType.CATEGORICAL),
       ColDef('week', DataType.DATE_TIME),
       ColDef('ln sales', DataType.NUMERIC, 
              ColType.OUTCOME),
       ColDef('featured', DataType.NUMERIC, 
              ColType.TREATMENT),
       ColDef('ln price', DataType.NUMERIC, 
              ColType.TREATMENT),
       ColDef('HVAL150', DataType.NUMERIC)]
schema = Schema(cols,
                time_colname='week',
                panel_colnames=['store id', 
                        'brand'])
dataset = EstimationDataSet(data, schema)
int_lvls = [['brand'], ['HVAL150']]
fbs = default_dynamic_featurizer.\
      get_featurizer(dataset.schema, 
                     min_lag=0, max_lag=7, 
                  exclude_dummies=['store'])
p2p = {}
p2p['dominicks']=['minute.maid','tropicana'],
p2p['tropicana']=['dominicks','minute.maid'],
p2p['minute.maid']=['tropicana','dominicks'],
tbs = [OwnVar('featured', 
              interaction_levels=int_lvls),
       OwnVar('log price', 
              interaction_levels=int_lvls),
       OwnVar('log price', lag=1),
       PToPVar('log price', 'brand', p2p)]
opts = DDMLOptions(min_lead=1, max_lead=3)
model = DynamicDML(dataset.schema,
                   baseline_model=LassoCV(),
                   causal_model=RidgeCV(),
                   options=opts,
                   feature_builders=fbs,
                   treatment_builders=tbs)
model.fit(dataset)
beta = model.get_coefficients()
se = model.get_standard_errors()
\end{lstlisting}

\section{Typical Analysis Process}

In this section, we briefly outline the general process of using the
Pricing Engine for causal estimation and show how it is implemented
in our OJ demand example.
\begin{enumerate}
\item Identify the main variables of interest (outcome and treatments),
time granularity (e.g. \texttt{week}), and unit (panel) identifying
variables (e.g. \texttt{region} $\times$ \texttt{channel} $\times$
\texttt{SKU}). This will be influenced both by data availability,
desired causal estimates, and useful variation in the data.\\
OJ: We use \texttt{ln sales} as our outcome and \texttt{ln price}
and \texttt{featured} as our treatments. Our data is weekly and individual
units are at the \texttt{store id} $\times$\texttt{ brand} level. 
\item Determine what information that decision makers used when modifying
the treatment in the past (e.g. competitor actions, product life-cycles,
and holidays). Divide this set into those that could independently
be affecting demand, which are potential confounders, and those that
do not. When in doubt, assume an element is potentially a confounder.
Only potential confounders should be included in the first-stage regression
of treatment.\\
OJ: We found that previous trends in the outcome and treatments were
used when setting new values of the treatments. We considered these
as potential confounders as they may be related to overall demand
changes.
\item Identify any additional variables that may be useful in predicting
the outcome variable. Any variables (excluding bad controls) that
improve prediction of the outcome can be usefully added to the first-stage
regression of outcomes. \\
OJ: We included the same features as with treatment.
\item Collect and prepare data. Make sure to collect data on all important
potential confounders.\\
OJ: Already done.
\item With these in place you should be able to use the Pricing Engine as
outlined above.\\
OJ: See section~\ref{sec:Implementing-Double-ML}.
\item Evaluate model results and potentially revise the model. (See section~\ref{subsec:Diagnostics})
\end{enumerate}

\appendix
\bibliographystyle{ieeetr}
\bibliography{user_guide}
 \onehalfspacing

\end{multicols*}
\end{document}